\documentclass[preprint2]{aastex}
\usepackage{graphicx,epsfig}
\newcommand{\Om}{\Omega_{\rm m}} 

\shorttitle{Systematics in Supernovae Data}
\shortauthors{Gupta et al.}

\begin{document}
\title {Direction Dependent Non-gaussianity in High-$z$ Supernova Data}

\author{Shashikant Gupta\altaffilmark{1} and Tarun Deep Saini\altaffilmark{2}}
\affil{Indian Institute of Science, Bangalore, Karnataka, India, 560 012}
\and
\author{Tanmoy Laskar\altaffilmark{3}}
\affil{St. Stephen's College, University of Delhi, Delhi, India 110 007}
\altaffiltext{1} {E-mail:shashikant@physics.iisc.ernet.in }
\altaffiltext{2}{E-mail:tarun@physics.iisc.ernet.in}
\altaffiltext{3}{E-mail:tanmoylaskar@gmail.com }
\begin{abstract}
  The most detailed constraints on the accelerating expansion of the
  universe and details of nature of dark energy are derived from the
  high redshift supernova data, assuming that the errors in the
  measurements are Gaussian in nature. There is a possibility that
  there are direction dependent systematics in the data, either due to
  uncorrected, known physical processes or because there are tiny
  departures from the cosmological principle, making the universe
  slightly anisotropic. To investigate this possibility we introduce a
  statistic based on the extreme value theory and apply it to the gold
  data set from \cite{rie04}. Our analysis indicates a systematic,
  direction dependent non-gaussianity at about one sigma level.
\end{abstract}


\section{Introduction}

During the last decade the possibility that the expansion of our
Universe is accelerating has been put on a firm footing. The combined
analysis of high redshift supernova data
\citep{per99,rie98,rie02,rie04}, along with observations of cosmic
microwave background \citep{ben02,pag06} or large scale structure
\citep{per02,teg04} indicates a spatially flat universe with low
matter density (around one-third of the critical density), the rest of
the closure density is believed to be in an unknown form generically
termed as the \emph{dark energy}. It is this component that
drives the late time acceleration of the expansion of the universe.

The simplest possibility, which fits the data well, is that the
acceleration is caused by the presence of a cosmological constant
term, called the $\Lambda$ term. When combined with the usual matter
term, the resultant model is referred to as the $\Lambda$CDM model. In
this model the Hubble parameter asymptotically approaches a constant
at late times, thus causing the universe to accelerate. There are
compelling theoretical reasons to believe that the dark energy density
may not be a strict constant. There are several physical models where
dark energy is generated dynamically from an evolving scalar field
(called quintessence) or even from alternate theories of gravity
\citep[see][for a recent review of models and methods of
reconstruction of cosmic history]{sah06}. Due to its simplicity, it
has become popular to phenomenologically model dark energy as an ideal
fluid with an equation of state given by $p=w\rho$, where $w$ is
allowed to be negative. In this model $w=-1$ gives the usual
cosmological constant. Models where $w$ is a constant or a simple
function of redshift have also been considered.

Cosmological data is rapidly approaching a quality where we can start
discriminating competing models of dark energy. The effect of tiny
departures from a strict cosmological constant on observations is
sufficiently small to render such analysis unreliable if the nature of
statistical noise in data is not well understood. Since the most
detailed constraints on dark energy are derived from the luminosity
distance to distant supernovae, we would like to be certain that their
statistics is well understood. The central limit theorem \citep{ken77}
ensures that, to a very good approximation, statistical noise due to a
large number of random influences can be treated as Gaussian. There
are several possible sources of non-gaussianity in the supernova data;
according to \citet{kol01} these could be 1) statistical scatter due
to location in the host galaxy and galaxy type 2) scatter due to dust
absorption in the host galaxy, inter galactic medium or in our own
galaxy or 3) due to lensing along the line of sight. Some of these
processes are corrected for in the data reduction. It is also possible
that the observed anisotropy is a result of collation of data of
disparate quality, perhaps due to differences in seeing condition or
in the data reduction process.

Modern cosmology is based on the Cosmological Principle (CP)
\citep{pee93}, which states that on the large scales the universe is
statistically homogenous and isotropic. Even if supernovae were
perfect indicator of distance, and if statistical noise in the
supernova distances is Gaussian, an anisotropic universe could
contribute a systematic, direction dependent modulation in the data.
Another possibility is that our galaxy could contain anisotropic, gray
dust patches, and since the dust correction depends on reddening in
the spectrum, this sort of dust could remain uncorrected. Above
arguments suggest that there is a strong case for investigating
direction dependence in the supernova data.

\citet{kol01} have investigated the possibility of detecting cosmic
anisotropy with $79$ high-$z$ supernovae obtained from \citet{rie98}
and \citet{per99}. In this paper we use a different statistic than
used by them. We use the gold data set containing $157$ supernovae
\citep{rie04} for our analysis. Although our methodology is derived
from testing for isotropy, we shall use the term non-gaussianity
interchangeably. Our results can be interpreted as a systematic
directional dependence in the data due to any of the above mentioned
possibilities.

The outline of this paper is as follows. In \S\ref{sec:method} we
describe our methodology in detail. We present our results in
\S\ref{sec:results} and our conclusions in \S\ref{sec:conclusions}.

\section{Methodology: Extreme Value Statistic} 
\label{sec:method}
Throughout our analysis we have assumed a flat FRW universe. Since
$\Lambda$CDM model fits the data quite well, we first obtain the best
fit model to the \emph{full gold data set} \citep{rie04} and calculate
the dispersion normalized residuals \mbox{ $\chi_i = [\mu_i -
  \mu^{\Lambda\rm CDM} (z_i;\Om)]/\sigma_{\mu}(z_i) $}, where the
distance modulus $\mu = 5\,\log(d_L/{\rm Mpc})+ 25$, the observed
values being $\mu_i$ for a supernova at redshift $z_i$, and
$\sigma_{\mu}(z_i)$ is the observed standard error. We shall consider
subsets of the full data set to construct our statistic. We define the
reduced $\chi^2$ in terms of $\chi_i$ as follows
\begin{equation}
\chi^2= \frac{1}{N_{\rm subset}}\sum_i \chi_i^2 \,\,,
\label{eq:chisq}
\end{equation}
where it should be noted that by `reduced' we do not mean `per degree
of freedom', since we \emph{do not fit} the model separately to the
subsets of the full data. Here $\chi^2$ is an indicator of the
statistical scatter of the subset away from the best fit $\Lambda$CDM
model.

If CP holds then the apparent magnitude of a supernova should not
depend upon the direction in which it is observed but only on the
cosmology. We divide the data into two hemispheres labeled by the
direction vector $\hat{n_i}$, and take the difference of the reduced
$\chi^2$ computed for the two hemispheres separately to obtain $\Delta
\chi_{\hat{n_i}}^2 =\chi^2_{\rm north} - \chi^2_{\rm south} $, where
we have defined 'north' as that hemisphere towards which the direction
vector $\hat{n_i}$ points. 

We take the absolute value of $\Delta \chi_{\hat{n_i}}^2$, since that
is the quantity which determines the plane across which data is
anisotropic, and then vary the direction $\hat{n}$ across the sky to
obtain the maximum absolute difference
\begin{equation}
 \Delta = {\rm max} \{| \Delta \chi_{\hat{n_i}}^2 |\}\,\,. 
\end{equation}
To interpret our results we need to know what values of $\Delta$ one
might expect. The distribution of supernovae is not uniform on the
sky, therefore, the number of supernovae in the two hemispheres for a
given direction varies with the direction $\hat{n}$. Therefore one
might expect the probability distribution function $P(\Delta)$ to be
extremely complicated, however, extreme value theory \citep{ken77}
shows that the distribution is, in fact, a simple, two parameter Gumbel
distribution, characteristic of extreme value distribution type~I:
\begin{equation}
P(\Delta) =\frac{1}{s} \exp \left[ -\frac{\Delta-m}{s}\right]\,\exp\left
  [-\exp\left(-\frac{\Delta-m}{s}\right)\right]\,\,, 
\end{equation} 
where the position parameter $m$ and the scale parameter $s$
completely determine the distribution. To quantify departures from
isotropy we need to know the theoretical distribution $P_{\rm
 theory}(\Delta)$. Since it is difficult to obtain it analytically, we
have calculated it numerically by simulating several sets of Gaussian
distributed $\chi_i$ on the gold set supernova positions and obtaining
$\Delta$ from each realization. We plot this distribution in
Fig~\ref{fig1} as the broken curve. We find that the
distribution closely resembles a Gumbel distribution.

If the data do have directional anisotropy then an independent
possible test for non-gaussianity is obtained by constructing the
bootstrap distribution $P_{\rm BS}(\Delta)$ in the following manner.
The observed $\chi_i$ are assumed to be drawn from some unknown,
direction dependent probability distribution. We shuffle the data
values $z_i$, $m(z_i)$ and $\sigma_m(z_i)$ over the supernovae
positions, thus destroying any directional alignment they might have
had due to anisotropy. We show in the next section that this gives us
yet another way of quantifying non-gaussianity.

\section{Results}
\label{sec:results}
\begin{figure}
\centering
\includegraphics[angle=0, width=0.45\textwidth]{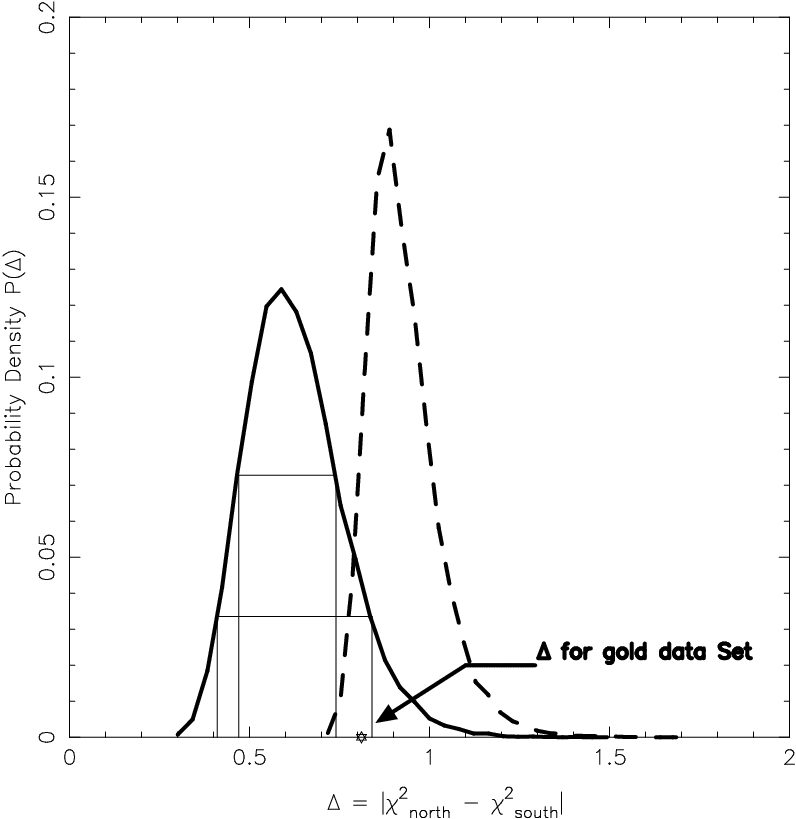}
\caption{
  The solid line shows the bootstrap probability distribution $P_{\rm
    BS}(\Delta)$ obtained by shuffling supernovae data on the sky
  while keeping the supernovae positions fixed. The broken line shows
  the theoretical distribution $P_{\rm theory}(\Delta)$ obtained by
  simulating Gaussian distributed $\chi_i$, as described in the text.
  The vertical lines show one and two sigma regions. The observed
  value for $\Delta$ for the gold data set is seen to be around one
  sigma away from either distribution.}
\label{fig1}
\end{figure}

Our main result is plotted in the Fig~\ref{fig1}. We find that
the theoretical distribution $P_{\rm theory}(\Delta)$ (broken line)
assuming Gaussian distributed $\chi_i$ indicates that our universe has
about one sigma smaller anisotropy than the mean of the distribution.
However, as mentioned in the last section, if the residuals $\mu_i$
are non-Gaussian then a more appropriate estimate of departure from
non-gaussianity would be the bootstrap distribution $P_{\rm
  BS}(\Delta)$. We have plotted $P_{\rm BS}(\Delta)$ obtained by
randomly shuffling $\chi_i$ on the given supernovae positions in the
same figure. We find that the observed value of $\Delta$ is more than
one sigma away from the mean of this distribution as well.

\begin{figure}

\centering
\includegraphics[angle=0, width=0.45\textwidth]{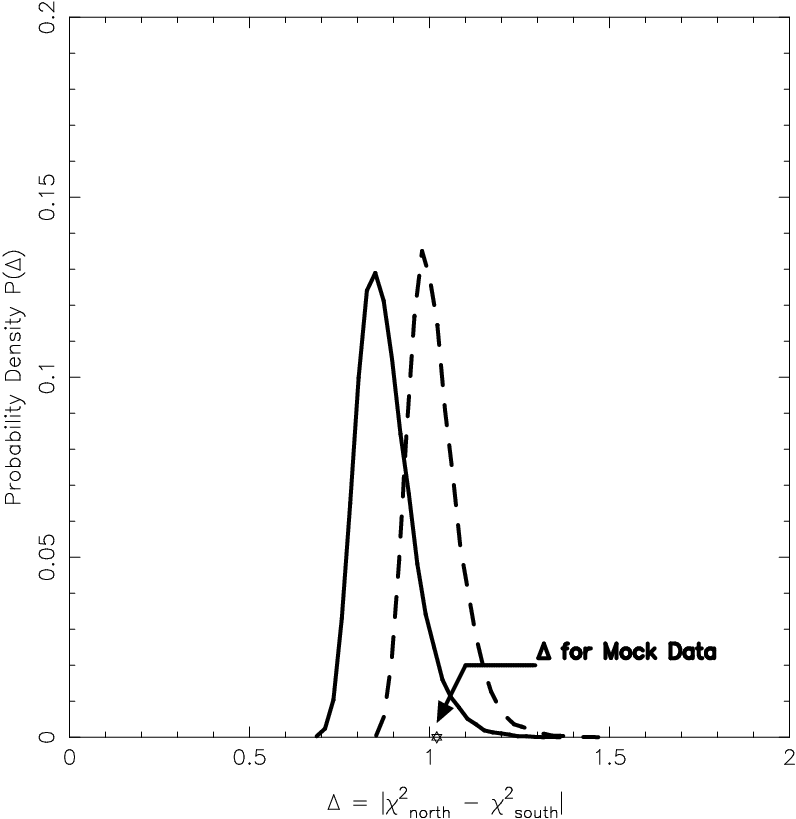}
\caption{
  We plot the result of a typical simulation where $157$ supernovae
  positions were generated randomly and populated with Gaussian noise.
  Similar to Fig~\ref{fig1} this plot shows the Gaussian vs. the
  bootstrap distribution. The simulated universe is seen to be
  consistent with the theoretical distribution, thus indicating that
  our statistic does not require a uniform sky coverage}
\label{fig2}
\end{figure}

One problem with the bootstrap distribution is that we expect it to be
shifted slightly to the left of the theoretical distribution. This is
due to the fact that theoretical distribution is obtained by assuming
$\chi_i$ to be Gaussian random variates with a zero mean and unit
variance. Therefore theoretical $\chi_i$s are unbounded. However, the
bootstrap distribution is obtained by shuffling through a
\emph{specific realization} of $\chi_i$, and they have a maximum value
such that $|\chi_i| < \chi_0$. Since the bootstrap realizations have
bounded $\chi_i$, they should produce, on the average, slightly
smaller values of $\Delta$ as compared to what one expects from a
Gaussian distributed $\chi_i$. For a large number of supernovae this
bias is expected to vanish. Since the gold data set contains only
$157$ supernovae, another concern is that our statistics may not be
reliable enough due to a lack of uniform sky coverage. We have made a
few checks to investigate these concerns: We simulate the same number
of supernovae as in the gold data set with randomly chosen positions
on the sky. We then process the simulated data in exactly the same way
as the actual data. A typical result is shown in Fig~\ref{fig2}. There
are a few things to be noted. 1) The simulated data gives a $\Delta$
that is consistent with the theoretical distribution, indicating that
a uniform sky coverage is not a strict requirement in this statistic.
2) The theoretical and bootstrap distributions for the simulated data
look identical in shape and 3) the peak of the bootstrap distribution
is shifted leftwards, as discussed above. This shift is on the order
of $10$ per cent. In Fig~\ref{fig1} we show that the bootstrap
distribution has a mean that is about $40$ per cent shifted from the
theoretical distribution, and has a \emph{different shape},
independently indicating non-gaussianity. The excess shift cannot be
reconciled with the theoretical distribution by a simple scaling of
the error bars. To produce a rightward shift we would need to increase
$\Delta$, which can be done by decreasing the error bars on supernovae
by a constant scale factor. However, this would also produce a
larger $\chi_{\nu}^2$ for the best fit $\Lambda$CDM model. The data actually
gives $\chi_{\nu}^2 = 1.14$, and scaling would make it larger, thus
making the primary fit worse.

\begin{figure}
\centering
\includegraphics[width=0.45\textwidth]{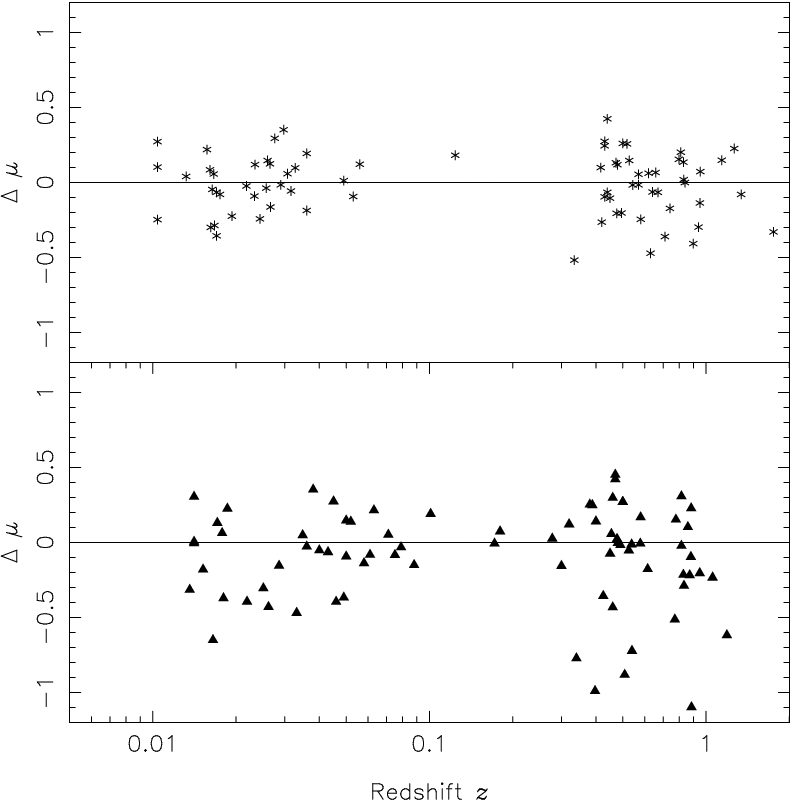}
\caption{
  The top panel shows the distribution of cool supernovae and the
  bottom one shows the hot supernovae. Horizontal line is the
  $\Lambda$CDM model in both panels. The total number of supernovae in
  the top panel is $74$ and the bottom panel $83$.}
\label{fig3}
\end{figure}

We find that the gold set is maximally asymmetric in the direction
$(l=100^{\circ},b=45^{\circ})$. We designate the two hemispheres as
`hot' or `cold' according to largeness or smallness of their reduced
$\chi^2$ (as given in Eq.~\ref{eq:chisq}) with respect to the best fit
$\Lambda$CDM model. In Fig~\ref{fig3} we have plotted the residuals
$\Delta \mu = \mu_{\rm data} - \mu_{\Lambda\rm CDM}$ for the hot and
cold subsets of supernovae. The zero line is the best fit $\Lambda$CDM
model. As expected, the cold supernovae show slightly smaller
dispersion compared to the hot ones. Figure~\ref{fig3} also shows
that the $\Lambda$CDM model seems to fit the hot and cold supernovae
equally well. This is borne out by a parameter estimation, which is
tabulated in Table~\ref{tbl-1}, where we find that the best fit
$\Lambda$CDM model for hot supernovae gives $\Om=0.30$ and the cold
ones give $\Om=0.31$, so the difference is only a few per cent.
However, the situation is not the same for model where we have assumed
a constant equation of state $p=w\rho$. We find that the model
parameters for the hot supernovae in this model are $\Om=0.51$ with $w
= -4.53$ and for the cold supernovae $\Om = 0.32$ and $w = -1.03$. The
value of the Hubble constant is relatively quite robust, showing that
most of the effect is coming from high-$z$ supernovae. The large
difference in the values for the constant $w$ model shows that the
level of non-gaussianity indicates that constraints on a more
complicated dark energy model are not as robust. Perhaps this explains
the intriguing result in \cite{ujj03}, that the data seems to fit a
$\Lambda$CDM model as well as a model with a strongly evolving dark
energy.

\section{Conclusions}
\label{sec:conclusions}

We have used the extreme value statistics on the gold data set and
shown the presence of non-Gaussian features in the data. We find that
there is a direction of maximal asymmetry in the data across which
data seems to imply different cosmological models for a constant
equation of state, although, the constraints on the $\Lambda$CDM model
are found to be robust. We have discussed how this could either imply
non-gaussianity in the data due to various possible physical processes
or due to anisotropy in the universe. Our results cannot be trivially
understood by scaling the error bars. Since this analysis has been
done within the framework of an FRW model, it is obviously very
difficult to quantify the precise meaning of this anisotropy. We have
discussed that we need to be very careful in interpreting dark energy
beyond the cosmological constant model since it is possible that
systematic noise may masquerade as evolving dark energy. Further work
is required to fully understand the statistic that we have introduced
in our analysis and will be discussed in greater detail in a future
publication.

\acknowledgments

Shashikant thanks CSIR for providing financial assistance for this
work. We thank Shiv Sethi for useful comments.

\def\etal{{\it et~al.\ }}
\def\apj{{Astroph.\@ J.\ }}
\def\mn{{Mon.\@ Not.\@ Roy.\@ Ast.\@ Soc.\ }}
\def\asta{{Astron.\@ Astrophys.\ }}
\def\aj{{Astron.\@ J.\ }}
\def\prl{{Phys.\@ Rev.\@ Lett.\ }}
\def\pd{{Phys.\@ Rev.\@ D\ }}
\def\nucp{{Nucl.\@ Phys.\ }}
\def\nat{{Nature\ }}
\def\plb {{Phys.\@ Lett.\@ B\ }}
\def \jetpl {JETP Lett.\ }

\begin{table}
\begin{center}
\caption{
  The model parameters for the hot and cold set are tabulated here.
  WCDM refers to a model where the dark energy has a constant equation
  of state.\label{tbl-1}} \bigskip
\begin{tabular}{crrrrrr}
\tableline\tableline
Model & Subset & $\Om$ & $ w $ & $H_0$ & $\chi_{\nu}^2$ \\
\tableline
$\Lambda$CDM & hot & 0.30 & -1 & 64.80  & 1.55 \\
$\Lambda$CDM & cold & 0.31 & -1 & 63.88  & 0.70 \\
WCDM & hot & 0.51 & -4.53 & 68.46  &  1.49 \\
WCDM & cold & 0.32 & -1.03 & 63.90  & 0.71 \\
\tableline
\end{tabular}
\end{center}
\end{table}
\end{document}